\documentclass{article}
\usepackage{iclr2027_conference,times}

\usepackage{amsmath,amsfonts,bm}

\def\eqref#1{equation~\ref{#1}}

\def\1{\bm{1}}

\def\vb{{\bm{b}}}

\def\ve{{\bm{e}}}

\def\vh{{\bm{h}}}

\def\vk{{\bm{k}}}

\def\vm{{\bm{m}}}

\def\vq{{\bm{q}}}

\def\vu{{\bm{u}}}

\def\vx{{\bm{x}}}
\def\vy{{\bm{y}}}
\def\vz{{\bm{z}}}

\DeclareMathAlphabet{\mathsfit}{\encodingdefault}{\sfdefault}{m}{sl}
\SetMathAlphabet{\mathsfit}{bold}{\encodingdefault}{\sfdefault}{bx}{n}

\usepackage{booktabs}
\usepackage{graphicx}
\usepackage{subcaption}
\newcommand{\std}[1]{\,\scalebox{0.85}{\normalfont\ensuremath{\pm #1}}}
\usepackage{hyperref}
\usepackage{placeins}
\usepackage{tabularx}
\usepackage{float}
\usepackage{tikz}
\usetikzlibrary{calc,positioning}
\usepackage{url}
\usepackage{multirow}
\usepackage{wrapfig}

\title{\raggedright\hyphenpenalty=10000\exhyphenpenalty=10000
VenusGenome: Genome-Aware Masked\\
Autoencoding for Microbial Genome\\
Representation Learning}

\author{
{\normalfont Liang Zhang$^{1,*}$ \quad Yukun Zhang$^{2,*}$ \quad Yang Tan$^{1}$ \quad Fan Jiang$^{3}$} \\[-1pt]
{\normalfont Bozitao Zhong$^{2}$ \quad Xiaoran Cheng$^{2}$ \quad Liang Hong$^{1,2,3,\dagger}$ \quad Mingchen Li$^{3,\dagger}$ \quad Yuanxi Yu$^{3,\dagger}$} \\[2pt]
{\normalfont\fontsize{8.5}{9.5}\selectfont $^{1}$School of Computer Science \quad $^{2}$School of Life Sciences and Biotechnology \quad $^{3}$Institute of Natural Sciences} \\[-1pt]
{\normalfont\fontsize{8.5}{9.5}\selectfont Shanghai Jiao Tong University, Shanghai, China} \\[-1pt]
{\normalfont\fontsize{8.5}{9.5}\selectfont $^{*}$Equal contribution. \quad $^{\dagger}$Corresponding authors. \quad
\texttt{\{hongl3liang,limc,y445980780\}@sjtu.edu.cn}}
}

\newcommand{\method}{VenusGenome}
\newcommand{\genes}{\mathcal{G}}
\newcommand{\mask}{\mathcal{M}}

\iclrfinalcopy

\begin{document}
\maketitle
\fancyhead{}

\begin{abstract}
Genomic-context models have shown that contextualizing protein embeddings can improve predictions of microbial function. We observe that a gene’s protein embedding is partially predictable from the embeddings of neighboring genes in the genome, motivating high-ratio masking that requires reconstruction from sparser genomic context. Genomic modeling must also distinguish biological relationships from assembly conventions, since the ordering of separate contigs does not imply adjacency and strand labels depend on the reference orientation. We therefore introduce VenusGenome, a gene-resolution masked autoencoder that combines visible-only encoding with genome-aware attention incorporating contig boundaries, relative strand orientations, and gene positions. We pretrain this model on a curated corpus of over 1.7 million microbial genome assemblies from isolate and metagenomic sources, using frozen protein embeddings as gene-level inputs. Across genome-level phenotype prediction and gene-level tasks, the resulting frozen representations achieve the highest mean performance among the evaluated models on nine of eleven primary metrics. An exploratory gene-input perturbation analysis further links representation sensitivity to experimentally measured fitness defects. Ablations support the contribution of genome-aware attention and show the benefits of masked autoencoding over BERT-style masked modeling, with further gains from higher masking ratios.
\end{abstract}


\section{Introduction}

Microbial function depends on both the proteins a genome encodes and the contexts in which their genes occur.
Gene order, orientation, and co-occurrence organize proteins into operons, biosynthetic regions, and other functional modules \citep{taboada2018operonmapper,hannigan2019deepbgc,carroll2021gecco}.
Learning representations that retain this organization could support predictions from individual-gene function to organismal phenotype.  

Protein language models provide transferable molecular representations, but independent protein encoding omits genomic organization \citep{rives2021biological,lin2023evolutionary}.
Nucleotide models retain sequence context at the cost of processing millions of bases per microbial assembly.
Gene-context models offer an intermediate resolution: gLM contextualizes protein embeddings in short metagenomic segments, and BacFormer extends ordered protein modeling to bacterial genomes \citep{hwang2024genomic,wiatrak2025bacformer}.
Their pretraining targets differ: gLM predicts continuous protein features, while BacFormer predicts discrete protein-family labels.
These models show that contextualizing protein embeddings can yield useful gene representations, while both use 15\% masking and retain masked positions in BERT-style encoders during pretraining.

However, sparse masking can leave substantial local context available for reconstruction.
In a preliminary analysis of our pretraining embeddings, a linear predictor using the two preceding and two following genes reduced
reconstruction error by 15.4\% relative to a constant mean predictor, compared with 8.9\% using four random genes from the same contig.
This result indicates that local neighborhoods provide readily accessible reconstruction cues beyond contig-level composition.
In vision, the availability of local reconstruction cues motivates high-ratio masking in masked autoencoders \citep{he2022masked}.
Although local predictability in protein embeddings differs from spatial
redundancy in images, our observation motivates higher masking ratios,
which reduce the availability of immediate neighbors and require
reconstruction from sparser genomic context.

Genomic organization introduces a complementary modeling challenge: distinguishing biological relationships from arbitrary assembly
conventions.
Existing models already incorporate structural information, including gene orientation in gLM and contig embeddings and rotary positional
encodings in BacFormer \citep{hwang2024genomic,wiatrak2025bacformer}.
However, the ordering of separate contigs does not establish genomic adjacency, and strand labels depend on the chosen reference orientation.
The same- or opposite-strand relationship between genes within a contig, in contrast, is preserved when that orientation is reversed.
These considerations motivate explicitly distinguishing within-contig interactions from cross-contig communication and encoding pairwise
strand relationships rather than relying solely on individual strand labels.

To address these limitations, we introduce \method{}, a microbial
genome model that learns contextualized gene and whole-genome
representations from frozen protein embeddings.
We represent each genome as an ordered sequence of gene-level
protein features and train an asymmetric masked autoencoder to
reconstruct missing features from genomic context.
Motivated by the observed predictability of protein embeddings from neighboring genes, we use 75\% masking to require reconstruction from sparser genomic context. The encoder processes only visible genes, while a lightweight decoder reconstructs the masked embeddings, separating contextual encoding from reconstruction and reducing encoder computation during pretraining.
To account for genomic organization, we combine contig-restricted
gene attention with relative strand biases and within-contig
positional information.
Global tokens mediate communication across contigs without
treating their concatenation as a continuous genomic neighborhood.
We pretrain \method{} on a curated corpus of over 1.7 million
microbial genome assemblies and evaluate the frozen encoder on
genome-level phenotype prediction and gene-level functional tasks.
We further examine whether sensitivity to gene-input perturbation
can identify genes with experimentally measured average fitness
defects.
Our main contributions are as follows:
\begin{itemize}
 \item We develop a genome-aware masked autoencoding framework that
combines visible-only contextual encoding with structured genomic
attention, and evaluate its main design choices through controlled
ablations. At 512M and matched 15\% masking, MAE reaches
    a zero-shot protein--protein interaction (PPI) genome-macro
    Spearman correlation of 0.282 versus 0.192 for BERT-style
    reconstruction; removing contig-aware attention from the
    75\%-masking model reduces it from 0.293 to 0.203.
    At this scale, increasing masking from 15\% to 75\% also
    improves correlation from 0.282 to 0.293 and reduces measured
    forward--backward time by 65.2\% for 4,096-gene inputs.

    \item We curate over 1.7 million isolate and metagenomic microbial
    genome assemblies into a unified gene-resolution pretraining
    corpus. Quality filtering, taxonomic standardization, and
    cross-source annotation integration preserve gene order, contig
    membership, and strand information, enabling genome-aware
    pretraining at scale.

    \item We evaluate one frozen encoder on six whole-genome phenotype
    tasks and five gene-level metrics spanning biosynthetic gene
    clusters (BGCs), essential genes, operons, and PPI.
    \method{} leads evaluated models on nine of eleven primary metrics,
    including zero-shot PPI ranking (0.300 versus 0.201 for BacFormer).
    An exploratory zero-shot analysis further links representation
    sensitivity to experimentally measured gene-disruption fitness,
    suggesting a potential route toward in silico functional screening
    in microbes.
\end{itemize}

\section{Related Work}

\paragraph{Biological sequence and genome-context models.}
Protein language models learn transferable molecular representations \citep{alley2019unified,elnaggar2022prottrans,rives2021biological,meier2021language,lin2023evolutionary}, with alignment-aware, generative, and structure-aware extensions \citep{rao2021msa,madani2023progen,su2024saprot}.
Nucleotide models learn from coding and non-coding sequences through masked, autoregressive, and long-context architectures \citep{ji2021dnabert,zhou2024dnabert2,dallatorre2023nucleotide,zvyagin2023genslm,nguyen2024hyenadna,schiff2024caduceus,nguyen2024evo,brixi2026evo2,fang2026genosm}.
At gene resolution, gLM contextualizes protein embeddings in metagenomic segments and predicts continuous protein features \citep{hwang2024genomic}, whereas BacFormer models ordered protein embeddings across bacterial genomes with protein-family prediction targets \citep{wiatrak2025bacformer}.
FGeneBERT combines protein-based gene representations with masked gene modeling and contrastive learning for metagenomics \citep{duan2024fgbert}.
Other approaches represent genomes through mixed coding and intergenic sequences, as in gLM2 \citep{cornman2025omg}, or protein sets, as in Protein Set Transformer \citep{martin2025proteinset}.

\begin{figure}[t!]
    \centering
    \begin{subfigure}{\textwidth}
    \centering
    \begin{tikzpicture}[inner sep=0pt,outer sep=0pt]
        \node (overview) {\includegraphics[width=0.89\textwidth]{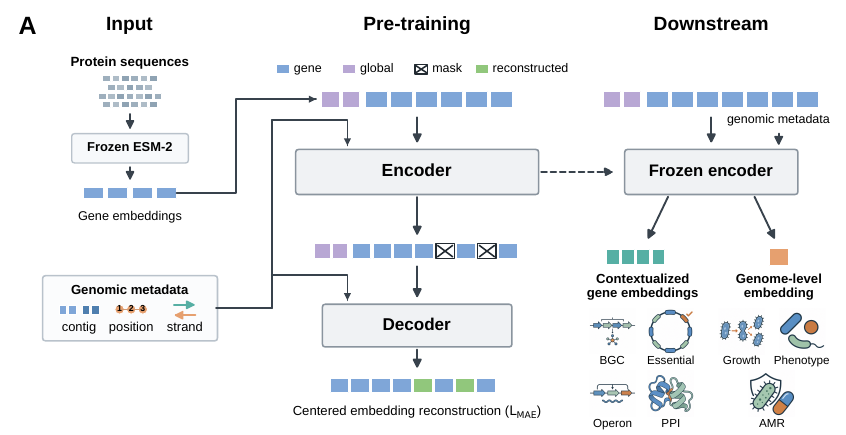}};
        \coordinate (overviewLabelTop) at ($(overview.north west)!0.030!(overview.north east)$);
        \coordinate (overviewLabelBottom) at ($(overview.south west)!0.030!(overview.south east)$);
        \coordinate (overviewLabel) at ($(overviewLabelTop)!0.050!(overviewLabelBottom)$);
        \node[fill=white,minimum width=1.35em,minimum height=1.35em] at (overviewLabel) {};
    \end{tikzpicture}
    \caption{Overall architecture}
    \label{fig:venusgenome-overview}
    \end{subfigure}
    \par\vspace{4pt}
    \begin{subfigure}{\textwidth}
    \centering
    \includegraphics[width=\textwidth]{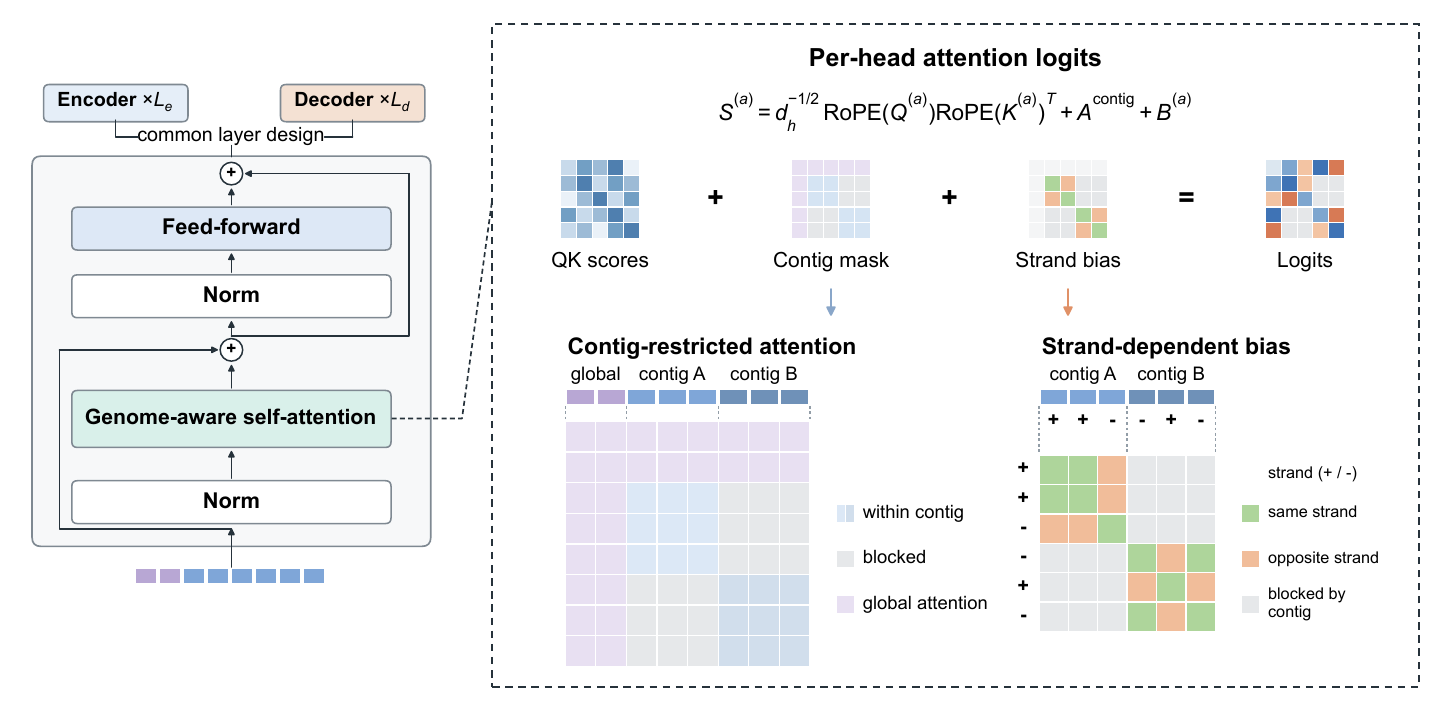}
    \caption{Genome-aware attention}
    \label{fig:venusgenome-attention}
    \end{subfigure}
    \caption{\method{} architecture and attention design. (a) Frozen protein embeddings and genomic metadata form gene inputs; a visible-only encoder and lightweight decoder perform masked reconstruction, and the frozen encoder supports downstream tasks. (b) Encoder and decoder use the same layer design with separate parameters. Contig masks and strand biases modify RoPE query--key logits; the expanded matrices show global connectivity and within-contig gene relationships.}
    \label{fig:venusgenome-model}
\end{figure}

\paragraph{Masked representation learning and structured attention.}
BERT and BART learn representations by predicting masked or corrupted inputs \citep{devlin2019bert,lewis2020bart}.
MAE instead encodes only visible inputs and uses a lightweight decoder for reconstruction, allowing high masking ratios to reduce encoder computation \citep{he2022masked}. MAE-LM extends masked autoencoding to language modeling and shows
  that excluding mask tokens from the encoder can improve the
  utilization of representation dimensions by real tokens
  \citep{meng2024representation}.
Complementing this architectural choice, MaskFeat predicts fixed features and I-JEPA predicts learned target representations, providing alternatives to raw-input reconstruction \citep{wei2022maskfeat,assran2023ijepa}.
Structured attention controls which inputs interact and how their relationships affect attention: BigBird combines local, random, and global connections \citep{zaheer2020bigbird}, while T5 introduces learned per-head relative-position biases \citep{raffel2020t5}.

\section{Method}

\method{} first maps frozen protein representations and assembly metadata to an ordered gene sequence, then learns contextualized states by reconstructing genes removed from a visible-only encoder.
The same encoder is used without masking for downstream gene- and genome-level representation extraction.
Figure~\ref{fig:venusgenome-overview} summarizes the overall workflow, and Figure~\ref{fig:venusgenome-attention} previews the genome-aware attention mechanisms described below.


\subsection{Inputs and Genome-aware Attention}

Let a microbial genome be an ordered sequence of $n$ protein-coding genes, $\genes=(g_1,\ldots,g_n)$.
Each gene $g_i$ has a 480-dimensional ESM2-35M embedding $\vx_i\in\mathbb{R}^{d}$ from the frozen protein language model \citep{lin2023evolutionary}, obtained by averaging residue states from its 12th (final) Transformer layer while excluding special and padding tokens, together with metadata $\vm_i=(c_i,p_i,s_i)$ denoting its contig, within-contig ordinal position, and strand.
We project the raw protein embedding as $\ve_i=W_x\vx_i+\vb_x$.
A per-dimension corpus mean $\boldsymbol{\mu}$, estimated over the pretraining protein embeddings, is used only to center the reconstruction targets; encoder inputs are not divided by a per-dimension standard deviation.

After selecting the visible genes, we prepend $G$ learned global tokens to the encoder sequence.
They attend bidirectionally to every visible gene and to one another, aggregating genome-wide information and relaying information between contigs that otherwise cannot communicate directly.
The resulting global states are projected into the decoder together with the visible-gene states, where they retain global connectivity to both visible and masked positions.

Metadata modifies self-attention directly.
We apply RoPE \citep{su2024roformer} using positions $\pi_i=p_i+\delta$ for gene tokens and $\pi_i=0$ for global tokens.
During pretraining, one offset $\delta$ is sampled uniformly from $\{0,\ldots,2047\}$ per input window and shared across its genes and both encoder and decoder; at inference, $\delta=0$.
For attention head $a$, let $\vq_i^{(a)}$ and $\vk_j^{(a)}$ denote query and key vectors of width $d_h$; the pre-softmax logit is
\begin{equation}
\label{eq:genome-attention}
    S^{(a)}_{ij}
    =
    \frac{\langle \mathrm{RoPE}_{\pi_i}(\vq_i^{(a)}),
    \mathrm{RoPE}_{\pi_j}(\vk_j^{(a)}) \rangle}{\sqrt{d_h}}
    + A^{\mathrm{contig}}_{ij}
    + B^{(a)}_{ij}.
\end{equation}
Here $A^{\mathrm{contig}}_{ij}=0$ when $i$ and $j$ belong to the same contig or either token is global, and $-\infty$ otherwise.
For gene pairs within the same contig, $B^{(a)}_{ij}$ equals the learned head-specific scalar $\alpha^{(a)}_{\mathrm{same}}$ or $\alpha^{(a)}_{\mathrm{diff}}$ according to whether their strands agree; it is zero for all other pairs.
The encoder and decoder learn separate strand-bias parameters, each shared across the layers of its respective module.
Figure~\ref{fig:venusgenome-attention} illustrates how these components enter each Transformer layer and determine gene connectivity.

\subsection{Masked Feature Reconstruction}

\method{} follows an asymmetric Transformer masked autoencoder
\citep{vaswani2017attention,he2022masked}.
Pretraining uses at most 4,096 genes per example, with a contiguous
gene window randomly cropped from longer assemblies.
We uniformly sample 75\% of valid genes to form the masked set
$\mask$, leaving the visible set
$\mathcal{V}=\{1,\ldots,n\}\setminus\mask$.
The encoder processes only the visible gene embeddings and global
tokens, preserving the genes' original within-contig positions.
The lightweight decoder projects the encoder states, inserts a learned
mask token at each masked gene position, and restores the original
gene order.
It retains the global tokens and uses the same positional, contig,
and strand mechanisms as the encoder, with separate learned parameters,
to produce gene states $\{\vu_i\}_{i=1}^{n}$.

We center the protein reconstruction targets to reduce the contribution
of shared embedding directions to cosine similarity, motivated by
anisotropy observed in language-model representations
\citep{ethayarajh2019contextual}.
In a diagnostic using ESM2 embeddings from 1,000 genomes, subtracting
the fixed pretraining mean reduces mean cosine similarity between
random cross-genome protein pairs from $0.830$ to $0.012$.
With targets $\vy_i=\vx_i-\boldsymbol{\mu}$ and predictions
$\hat{\vy}_i=r_\phi(\vu_i)$, we minimize reconstruction loss only
at masked gene positions:
\begin{equation}
\label{eq:masked-reconstruction}
  \mathcal{L}_{\mathrm{MGM}}
  = \frac{1}{|\mask|}\sum_{i\in\mask}
  \left[
  \frac{\lambda_{\mathrm{mse}}}{d}
  \|\hat{\vy}_i-\vy_i\|_2^2
  + \lambda_{\mathrm{cos}}
  \bigl(1-\cos(\hat{\vy}_i,\vy_i)\bigr)
  \right].
\end{equation}
We set $\lambda_{\mathrm{mse}}=0.3$ and
$\lambda_{\mathrm{cos}}=0.7$, combining Euclidean reconstruction
with angular alignment.
The canonical model pairs a 44-layer encoder with an 8-layer decoder;
Appendix~\ref{app:pretraining-hyperparameters} provides full
architecture and optimization settings.

\subsection{Representation Extraction}

For downstream extraction, we set the mask ratio to zero, discard the decoder, and encode every valid gene.
Gene-level tasks use the resulting contextualized state $\vh_i$ directly; genome-level tasks use either the mean of all gene states or the mean of the final encoder global states,
\begin{equation}
\label{eq:genome-readouts}
    \vz_{\mathrm{mean}}=\frac{1}{n}\sum_{i=1}^{n}\vh_i,
    \qquad
    \vz_{\mathrm{global}}=\frac{1}{G}\sum_{j=1}^{G}\vh^{\mathrm{global}}_j.
\end{equation}
For assemblies longer than 4,096 genes, we use overlapping 4,096-gene windows with 512-gene overlap, average repeated gene states across windows, and average the window-level global readouts.
Windows are encoded independently, so stitching covers every gene but does not introduce direct attention across windows.
Main genome results use $\vz_{\mathrm{mean}}$ from the stitched gene states.
Appendix~\ref{app:genome-embedding-geometry} compares both readouts on the same species-disjoint benchmarks.

\FloatBarrier
\section{Experiments}

We evaluate \method{} across genome- and gene-level benchmarks,
covering six whole-genome phenotype tasks, BGC detection and category
prediction, essential-gene prediction, and zero-shot operon and PPI
ranking. We also assess zero-shot gene fitness prediction using experimental measurements from Fitness Browser.

\subsection{Experimental Setup}

\paragraph{Pretraining setup.}
We assemble a pretraining corpus of over 1.7 million microbial
genomes from NCBI, proGenomes v4, MGnify, and SPIRE
\citep{sayers2022ncbi,fullam2026progenomes4,gurbich2023mgnifygenomes,schmidt2024spire}.
For isolate genomes, we limit overrepresented species and apply
completeness and contamination filters, retaining 75,390
high-quality assemblies.
Unless otherwise specified, \method{} denotes the 1B encoder
pretrained for 40k steps with 75\% gene masking and contexts of
up to 4,096 genes; 85M and 512M variants are used for scaling analysis.
Appendices~\ref{app:datasets} and~\ref{app:pretraining-hyperparameters}
detail corpus construction and pretraining settings.

\paragraph{Downstream evaluation.}
We compare ESM2 protein embeddings \citep{lin2023evolutionary}, gLM and BacFormer gene-context embeddings \citep{hwang2024genomic,wiatrak2025bacformer}, and Nucleotide Transformer v2 (NT-v2) and Evo2 embeddings \citep{dallatorre2023nucleotide,brixi2026evo2}.
For genome-level tasks, both nucleotide baselines aggregate
features extracted from DNA windows across each assembly,
including coding and intergenic regions; \method{} uses mean pooling
of contextualized gene states.
All representation encoders remain frozen.
The six supervised genome-level tasks use species-disjoint splits,
while BGC and essential-gene prediction use genome-disjoint splits.
Within each task, we use the same one-hidden-layer MLP probe,
optimization settings, fixed data split, and validation-based
checkpoint selection across representations, adjusting input
dimensions to the embedding size.
Supervised results are averaged over five probe-training seeds;
operon, PPI, and gene fitness evaluations are zero-shot.
Appendices~\ref{app:downstream-protocols}--\ref{app:pretraining-hyperparameters}
provide evaluation and extraction details.

\begin{table}[h!]
\caption{Genome-level supervised results under strict species-disjoint splits.}
\label{tab:genome-supervised-results}
\begin{center}
\fontsize{8.5}{10}\selectfont
\setlength{\tabcolsep}{1.5pt}
\medmuskip=1mu
\renewcommand{\arraystretch}{1.12}
\begin{tabular*}{\linewidth}{@{\extracolsep{\fill}}lcccccc@{}}
\toprule
\multirow{3}{*}[-0.6ex]{Model} & \multicolumn{3}{c}{\textit{(a) Regression}} & \multicolumn{3}{c}{\textit{(b) Classification}} \\
\cmidrule(lr){2-4}\cmidrule(lr){5-7}
& Temperature & Doubling & MIC & AMR & Binary phen. & Mixed phen. \\
& Pearson & Pearson & Macro Pearson & Macro AUROC & Macro AUROC & Macro F1 \\
\midrule
ESM2-35M & $0.845\std{0.005}$ & $0.212\std{0.016}$ & $0.110\std{0.041}$ & $0.520\std{0.018}$ & $0.771\std{0.004}$ & $0.420\std{0.025}$ \\
ESM2-650M & $\mathbf{0.877}\std{0.002}$ & $0.217\std{0.010}$ & $0.197\std{0.038}$ & $0.574\std{0.010}$ & $0.783\std{0.003}$ & $0.389\std{0.025}$ \\
gLM & $0.848\std{0.004}$ & $0.072\std{0.146}$ & $0.175\std{0.019}$ & $0.560\std{0.022}$ & $0.780\std{0.006}$ & $0.410\std{0.022}$ \\
BacFormer & $0.824\std{0.001}$ & $0.185\std{0.011}$ & $0.254\std{0.015}$ & $0.614\std{0.018}$ & $0.788\std{0.001}$ & $0.421\std{0.023}$ \\
NT-v2 & $0.755\std{0.005}$ & $0.164\std{0.047}$ & $0.153\std{0.024}$ & $0.633\std{0.026}$ & $0.756\std{0.008}$ & $0.367\std{0.011}$ \\
Evo2 & $0.476\std{0.009}$ & $0.254\std{0.005}$ & $0.152\std{0.021}$ & $0.555\std{0.063}$ & $0.687\std{0.006}$ & $0.266\std{0.014}$ \\
\addlinespace[2.5pt]
\method{} & $0.862\std{0.001}$ & $\mathbf{0.336}\std{0.011}$ & $\mathbf{0.270}\std{0.016}$ & $\mathbf{0.682}\std{0.034}$ & $\mathbf{0.794}\std{0.004}$ & $\mathbf{0.457}\std{0.027}$ \\
\bottomrule
\end{tabular*}
\renewcommand{\arraystretch}{1.0}
\end{center}
\end{table}

\begin{table}[h!]
\caption{Gene-level results.}
\label{tab:gene-region-pair-results}
\begin{center}
\scriptsize
\setlength{\tabcolsep}{3.0pt}
\renewcommand{\arraystretch}{1.16}
\resizebox{\linewidth}{!}{
\begin{tabular}{lccccc}
\toprule
\multirow{2}{*}[-0.6ex]{Model} & \multicolumn{2}{c}{BGC} & Essential & Operon & STRINGDB PPI \\
\cmidrule(lr){2-3}\cmidrule(lr){4-4}\cmidrule(lr){5-5}\cmidrule(lr){6-6}
& Det. AUPRC & Cat. macro F1 & AUPRC & AUPRC & Genome $\rho$ \\
\midrule
ESM2-35M  & $0.358\std{0.001}$ & $0.400\std{0.003}$ & $0.695\std{0.013}$ & $0.287$ & $0.102$ \\
ESM2-650M  & $0.404\std{0.001}$ & $0.450\std{0.003}$ & $\mathbf{0.714}\std{0.009}$ & $0.289$ & $0.046$ \\
gLM  & $0.525\std{0.002}$ & $0.528\std{0.006}$ & $0.504\std{0.006}$ & $0.275$ & $0.084$ \\
BacFormer  & $0.536\std{0.001}$ & $0.537\std{0.008}$ & $0.637\std{0.008}$ & $0.349$ & $0.201$ \\
NT-v2  & $0.235\std{0.001}$ & $0.253\std{0.003}$ & $0.330\std{0.008}$ & $0.308$ & $0.083$ \\
Evo2  & $0.392\std{0.003}$ & $0.334\std{0.008}$ & $0.326\std{0.009}$ & $0.335$ & $0.170$ \\
\addlinespace[2.5pt]
\method{} & $\mathbf{0.667}\std{0.003}$ & $\mathbf{0.595}\std{0.003}$ & $\mathbf{0.714}\std{0.031}$ & $\mathbf{0.388}$ & $\mathbf{0.300}$ \\
\bottomrule
\end{tabular}
}
\renewcommand{\arraystretch}{1.0}
\end{center}
\end{table}
\FloatBarrier

\subsection{Genome-Level Evaluation}

We evaluate genome-level representations on six supervised tasks covering microbial growth traits (optimal growth temperature and
population doubling time), antimicrobial response (MIC regression
and multilabel AMR classification), and binary and categorical
microbial phenotypes.
To assess transfer to unseen species, we use species-disjoint
training, validation, and test sets defined by NCBI species taxon
identifiers.
Table~\ref{tab:genome-supervised-results} summarizes the results;
task-specific metrics and evaluation protocols are detailed in
Appendix~\ref{app:downstream-protocols}.

\method{} achieves the highest mean performance on five of six
tasks under species-disjoint evaluation.
The strongest competing baseline varies across traits:
ESM2-650M leads on temperature, Evo2 on doubling time,
NT-v2 on AMR, and BacFormer on MIC and both phenotype tasks.
The gains therefore extend across traits favored by different
representation models, with doubling-time Pearson correlation
increasing from $0.254$ for Evo2 to $0.336$.
\method{} also consistently outperforms mean pooling of ESM2-35M
embeddings, which serve as its frozen inputs, showing that
genome-level pretraining improves transfer beyond simply
aggregating individual-protein representations.

\subsection{Gene-Level Evaluation}

We evaluate gene representations on supervised BGC and essential-gene prediction, together with zero-shot operon and PPI ranking. Zero-shot operon and PPI ranking further tests whether the learned representation geometry reflects functional relationships between genes, without a trained prediction head. Supervised evaluations use genome-disjoint splits; BGC detection includes all labeled genes, whereas category prediction is restricted to BGC-positive genes. Both zero-shot tasks use gene-embedding cosine similarity, with PPI performance measured by its Spearman correlation with STRINGDB confidence scores \citep{szklarczyk2023string} within each genome and averaged equally across genomes. Table~\ref{tab:gene-region-pair-results} summarizes the results; detailed evaluation protocols are provided in Appendix~\ref{app:downstream-protocols}.

\begin{figure}[!t]
    \centering
    \includegraphics[width=\textwidth]{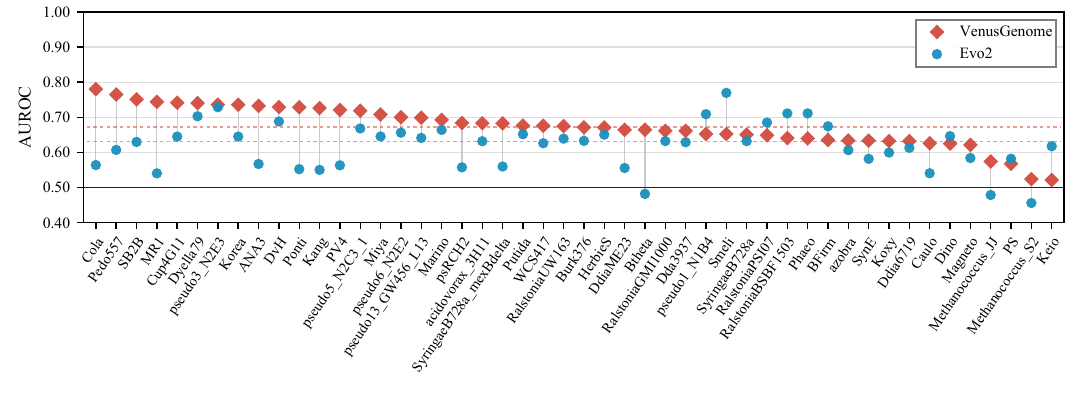}
    \caption{Zero-shot gene fitness evaluation on Fitness Browser 2024. Positive genes have mean experimental fitness $\leq -1$. Points show within-organism AUROC for all 48 organisms, ordered by \method{}; gray segments pair models and colored dashed lines mark medians. The black line marks random ranking (AUROC $=0.50$).}
    \label{fig:fitness-browser-zero-shot}
\end{figure}

Table~\ref{tab:gene-region-pair-results} shows that \method{} leads
four of five metrics and performs comparably to ESM2-650M on
essential-gene prediction.
It improves on ESM2-35M, which supplies its frozen input embeddings,
across all five metrics, demonstrating the value of additional
genome-level pretraining.
Notably, increasing ESM2 from 35M to 650M improves supervised
performance but does not consistently improve zero-shot ranking.
This distinction highlights that features useful to a trained
predictor need not encode functional relationships directly
through cosine similarity.
\method{} improves both supervised prediction and zero-shot ranking,
with PPI genome-macro Spearman correlation reaching $0.300$,
compared with $0.201$ for BacFormer, the strongest competing baseline.
Section~\ref{sec:inference-context}  further examines how extraction
context length affects BGC prediction and zero-shot PPI ranking.


\subsection{Zero-Shot Gene Fitness Evaluation}
\label{sec:gene-effect-evaluation}

We ask whether sensitivity to gene-input perturbation can identify genes associated with negative average fitness effects
across assayed experiments.
Fitness Browser 2024 provides gene fitness measurements from 7,552 pooled transposon-insertion experiments across 48 organisms
\citep{price2024fitnessbrowser}.
Fitness reflects normalized log$_2$ changes in mutant abundance, with negative values indicating a competitive disadvantage
under the tested conditions. We hypothesize that genes with greater influence on neighboring representations, or higher sequence likelihood under Evo2,
are more likely to be functionally important, such that their
disruption leads to larger fitness defects.

Without fitting a downstream predictor, we replace each target gene's protein embedding with its organism's mean embedding and
measure the change in the mean contextual representation of nearby genes.
We label genes with mean experimental fitness $\leq -1$ as positive and all remaining evaluated genes as negative, and compute
AUROC separately within each organism.
\method{} achieves a median AUROC of $0.673$, compared with $0.631$ for an exploratory Evo2 baseline using standalone CDS
likelihood, with higher AUROC in 39 of 48 organisms (Figure~\ref{fig:fitness-browser-zero-shot}).
These results suggest that local representation sensitivity carries information about experimentally measured average fitness
defects.
Appendix~\ref{app:gene-effect} details the experimental targets, sensitivity scores, and evaluation protocol.


\section{Analysis}


\subsection{Pretraining Design}

\begin{figure}[!t]
    \centering
    \includegraphics[width=\linewidth]{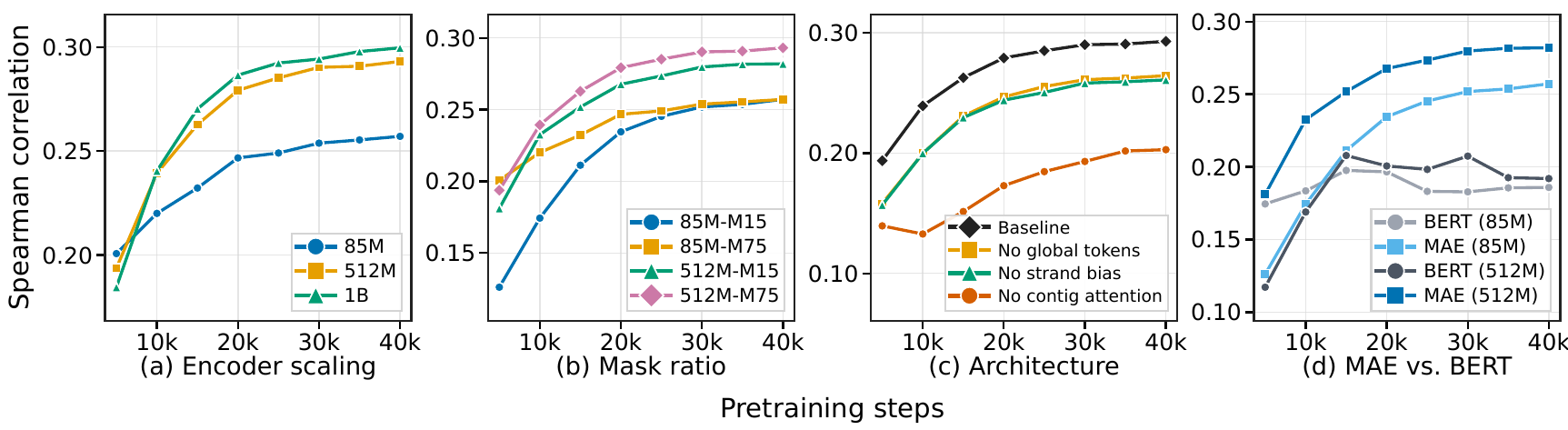}
    \caption{PPI scaling and design analysis. All panels report zero-shot genome-macro Spearman correlation on STRINGDB. (a) Encoder scaling at 75\% masking. (b) Mask-ratio comparison. (c) Architectural ablations for the 512M model. (d) MAE versus BERT-style full-sequence reconstruction at 15\% masking. Curves show single pretraining runs.}
    \label{fig:ppi-ablation-summary}
\end{figure}

We analyze encoder scaling and conduct ablation studies on masking
ratio, genome-aware attention, and pretraining formulation.
We use zero-shot PPI ranking to test whether functional relationships
between genes are reflected in the learned representation geometry.
By comparing embedding cosine similarity with STRINGDB confidence
scores, this evaluation tracks changes across checkpoints without
fitting a downstream predictor.
Figure~\ref{fig:ppi-ablation-summary} summarizes these comparisons.

Increasing encoder size improves zero-shot PPI ranking under
75\% masking, although the gain from 512M to 1B is smaller than
that from 85M to 512M (Figure~\ref{fig:ppi-ablation-summary}a).
Higher masking does not uniformly improve ranking: moving from
15\% to 75\% improves the 512M model but leaves the 85M result
nearly unchanged (Figure~\ref{fig:ppi-ablation-summary}b).
For the benchmarked 4,096-gene configuration, 75\% masking reduces
measured forward--backward time by $65.2\%$ and peak memory by
$51.3\%$ relative to 15\% masking
(Appendix~\ref{app:mask-efficiency}).
These results complement the local-predictability analysis in
the Introduction: although neighboring genes provide readily
accessible reconstruction cues, reducing their availability
during pretraining preserves or improves PPI ranking in the
evaluated configurations while substantially reducing computation.

\begin{wrapfigure}[15]{r}{0.43\textwidth}
\centering
\includegraphics[width=\linewidth]{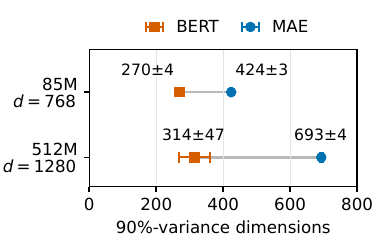}
\captionsetup{font=small,skip=3pt}
\caption{Centered gene-representation dimensionality
(mean $\pm$ SD over three genome-sampling seeds).}
\label{fig:representation-dimensionality}
\end{wrapfigure}

Architectural ablations show that removing contig-aware attention
causes the largest PPI degradation among the tested components
(Figure~\ref{fig:ppi-ablation-summary}c).
At 512M, removing it reduces correlation from $0.293$ to $0.203$;
removing global tokens or strand bias produces smaller declines.
The pretraining comparison further shows that MAE outperforms
BERT-style full-sequence reconstruction at both evaluated scales
under 15\% masking (Figure~\ref{fig:ppi-ablation-summary}d).
Both formulations reconstruct the same continuous targets with
the same loss, but MAE separates visible-gene encoding from
masked-gene reconstruction through a lightweight decoder. Appendix~\ref{app:pretraining-loss} provides complementary
pretraining-loss diagnostics.

We next examine whether MAE and BERT-style models differ in how
broadly variation is distributed across their gene-representation
spaces, following diagnostics in MAE-LM
\citep{meng2024representation}.
We compare 85M and 512M checkpoints pretrained for 40k steps
with 15\% masking.
For each of three genome-sampling seeds, we sample 256
isolate-source pretraining genomes and extract final normalized
gene representations from the same unmasked windows of up to
512 genes.
We select up to 64 identical gene positions per window for both
models, excluding global tokens.
For each model, we subtract the mean of all sampled gene
representations and use PCA to count the components needed to
explain 90\% of the variance.
MAE requires more components than BERT at both scales
(Figure~\ref{fig:representation-dimensionality}), showing that
variation among its gene representations is less concentrated
in a small number of directions.

\FloatBarrier

\subsection{Context at Inference}
\label{sec:inference-context}

We next ask how much genomic context is needed when extracting
representations from a fixed pretrained model.
Using the same 1B mask-75\% checkpoint, we vary the extraction
window from 9 to 4,096 genes and evaluate BGC prediction and
zero-shot PPI ranking.
The \emph{single gene + global} control retains the pretrained
encoder and all 64 global tokens but encodes each gene independently,
removing access to other gene inputs.
For a controlled comparison across context lengths, all BGC settings
use the same fixed genome-disjoint split and five probe seeds as
Table~\ref{tab:gene-region-pair-results}.
PPI uses the same test pairs and genome-macro Spearman correlation
as the main evaluation.

\begin{figure}[!htbp]
  \centering
  \includegraphics[width=\linewidth]{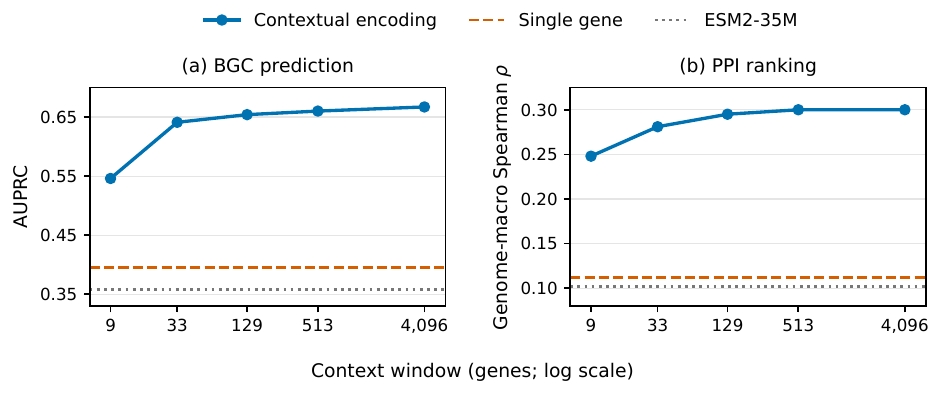}
  \caption{Effect of extraction context length with a fixed pretrained
  encoder. BGC reports mean AUPRC over five probe seeds on a fixed
  genome-disjoint split; PPI reports genome-macro Spearman correlation.
  Horizontal lines show the single-gene control and ESM2-35M baseline.
  Context length is plotted on a logarithmic axis.}
  \label{fig:context-decoupling}
\end{figure}

  Figure~\ref{fig:context-decoupling} shows that both tasks benefit
  from additional gene context.
  A 33-gene window recovers approximately 90\% of the improvement
  over the single-gene control observed at the maximum window size
  for both metrics.
  The benefit of further expansion differs between tasks:
  BGC AUPRC continues to increase, from $0.641$ at 33 genes
  to $0.667$ at 4,096 genes, whereas PPI correlation shows
  diminishing gains and is nearly unchanged beyond 513 genes.
  Thus, local neighborhoods account for much of the observed
  context benefit, while larger windows provide additional,
  more modest gains for BGC prediction.

\FloatBarrier

\section{Conclusion}



We introduce \method{}, a microbial genome model that combines high-ratio masked autoencoding with genome-aware attention to learn contextualized gene and whole-genome representations from protein embeddings. Pretrained on a curated corpus of over 1.7 million microbial genome assemblies, \method{} achieves the highest mean performance among the evaluated models on nine of eleven primary metrics using frozen representations. These results support its use as a shared representation model for microbial functional prediction at both gene and genome levels.

This study has several limitations that motivate further work. First, the use of frozen protein embeddings constrains contextual learning to the information retained by the protein encoder. Stronger protein representations, joint training, or complementary nucleotide information could expand the information available to the model. Second, our exploratory gene-fitness analysis evaluates average effects across assayed experiments and does not resolve condition-specific responses. Integrating genomic representations with environmental context and condition-specific perturbation measurements could extend this framework toward predicting how gene function and perturbation effects vary across conditions.

\clearpage

\section*{Code Availability}
The GitHub repository and Hugging Face model links will be made public
when the paper is published online.


\bibliography{iclr2027_conference}
\bibliographystyle{iclr2027_conference}

\clearpage
\appendix

\section{Experimental Details}
\label{app:experimental-details}

\subsection{Datasets}
\label{app:datasets}

\begin{figure}[!htbp]
    \centering
    \includegraphics[width=0.66\linewidth]{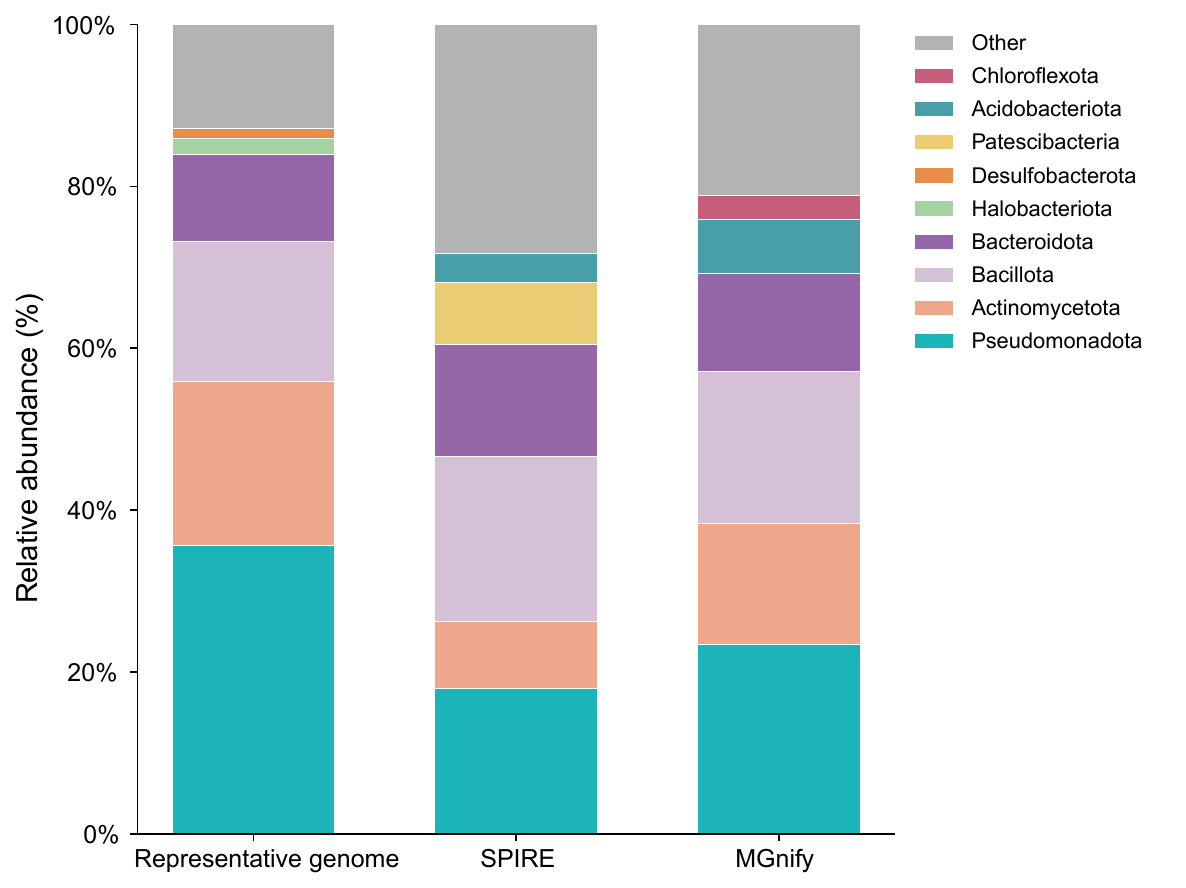}
    \caption{Phylum-level composition of the representative high-quality isolate collection, SPIRE, and MGnify at the species-cluster level. Each stacked bar reports the relative abundance of non-redundant species clusters within one source. The legend is the union of the six most abundant phyla selected separately in each source; phyla outside a source-specific top six are grouped as \emph{Other} for that source. Each species cluster is counted once within a source, and phylum labels are harmonized across the source taxonomies.}
    \label{fig:pretraining-phylum-composition}
\end{figure}

The pretraining corpus combines isolate genomes from NCBI and proGenomes v4 \citep{sayers2022ncbi,fullam2026progenomes4} with metagenome-assembled genomes (MAGs) from MGnify and SPIRE \citep{richardson2023mgnify,gurbich2023mgnifygenomes,schmidt2024spire}.
Isolate taxonomy was standardized against LPSN \citep{parte2020lpsn}, and the collection was capped at five genomes per species.
Only the 75,390 high-quality isolate genomes passing CheckM \citep{parks2015checkm} thresholds of $\geq90\%$ completeness and $\leq5\%$ contamination were included in the pretraining corpus.
Genes were predicted with Prodigal \citep{hyatt2010prodigal} for isolates and taken from source annotations for MAGs; gene order, contig, and strand were retained.
The MAG collection contains more than 1.67 million genomes, with more than 510k from MGnify and 1.16 million from SPIRE.
Figure~\ref{fig:pretraining-phylum-composition} compares their phylum-level composition after species-level dereplication.

\FloatBarrier

Genome-level benchmarks cover six tasks: optimal growth temperature, doubling time, binary antimicrobial resistance, minimum inhibitory concentration (MIC), binary phenotypes, and mixed phenotypes, using data derived from BacDive, BacBench, and BacFormer \citep{reimer2022bacdive,wiatrak2025bacbench,wiatrak2025bacformer}.
All six main genome-level results use species-disjoint splits based on NCBI species taxon identifiers, with no species shared across train, validation, and test.
Positive BGC labels were assigned to genes matched to known clusters by antiSMASH KnownClusterBlast against MIBiG, while unmatched genes in the benchmark genomes were treated as negatives\citep{terlouw2023mibig, blin2023antismash}.
Essential-gene, operon, and STRINGDB PPI benchmarks follow BacBench \citep{wiatrak2025bacbench,szklarczyk2023string}.
The separate Fitness Browser analysis uses the February 2024 release of pooled randomly barcoded transposon sequencing (RB-TnSeq) experiments, which measure the fitness effects of gene disruption \citep{price2024fitnessbrowser}.
This analysis includes 179,946 genes with fitness records across 48 organisms.
Table~\ref{tab:dataset-overview} summarizes the downstream datasets after filtering and split construction.

\begin{table}[h!]
\caption{Downstream datasets and probe settings.
Split sizes give train/validation/test genome counts;
zero-shot tasks report evaluation genomes in the test position and have no probe.}
\label{tab:dataset-overview}
\begin{center}
\footnotesize
\setlength{\tabcolsep}{3.0pt}
\renewcommand{\arraystretch}{1.25}
\renewcommand{\tabularxcolumn}[1]{m{#1}}
\begin{tabularx}{\textwidth}{@{}>{\raggedright\arraybackslash}m{0.28\textwidth}>{\raggedright\arraybackslash}m{0.20\textwidth}>{\centering\arraybackslash}m{0.30\textwidth}>{\raggedright\arraybackslash}X@{}}
\toprule
Dataset & Metric & Split sizes & Loss \\
\midrule
\multicolumn{4}{@{}l}{\textit{Genome-level prediction}} \\
\addlinespace[2pt]
Optimal growth temperature & Pearson & 3,713 / 795 / 795 & MSE \\
Doubling time & Pearson & 1,249 / 267 / 266 & MSE \\
Antibiotic resistance & Macro AUROC & 17,997 / 4,291 / 3,519 & BCE \\
MIC & Macro Pearson & 17,997 / 4,291 / 3,519 & MSE \\
Binary phenotypes & Macro AUROC & 17,080 / 3,660 / 3,660 & BCE \\
Mixed phenotypes & Macro F1 & 17,080 / 3,660 / 3,660 & MSE $+$ CE \\
\addlinespace[5pt]
\multicolumn{4}{@{}l}{\textit{Gene-level prediction}} \\
\addlinespace[2pt]
BGC detection / category & \shortstack[l]{AUPRC /\\macro F1} & 555 / 61 / 153 & \shortstack[l]{Weighted BCE\\$+$ CE} \\
\addlinespace[2pt]
Essential genes & AUPRC & 30 / 11 / 10 & BCE \\
Fitness Browser 2024 & AUROC & --- / --- / 48 & --- \\
Operons & AUPRC & --- / --- / 11 & --- \\
PPI & Spearman $\rho$ & --- / --- / 53 & --- \\
\bottomrule
\end{tabularx}
\renewcommand{\arraystretch}{1.0}
\end{center}
\end{table}
\FloatBarrier

\subsection{Evaluation Protocols}
\label{app:downstream-protocols}

All supervised evaluations use precomputed embeddings with frozen
representation encoders.
We use an MLP probe consisting of
LayerNorm--Linear--GELU--Dropout--Linear, with dropout 0.1.
Within each task, all representations use the same probe architecture
and training hyperparameters, with input dimensions adjusted to the
embedding size.
Probes are trained with AdamW at a learning rate of $10^{-3}$
and weight decay of 0.01 for up to 100 epochs, with early stopping
after 15 epochs without improvement in the task-specific validation
criterion.
For multi-output genome tasks, we average the applicable MSE,
binary cross-entropy, or categorical cross-entropy losses over
observed targets.
Table~\ref{tab:dataset-overview}  summarizes task-specific settings.

Zero-shot operon evaluation ranks gene pairs by embedding cosine
similarity and reports AUPRC.
For PPI, we treat each protein pair as unordered and count it once,
compute the Spearman correlation between embedding cosine similarity
and STRINGDB confidence scores within each genome, and report
the unweighted mean across genomes.
Appendix~\ref{app:gene-effect} provides the protocol for zero-shot
gene fitness evaluation.


\subsection{Gene Fitness Evaluation Details}
\label{app:gene-effect}

We use the February 2024 Fitness Browser release, covering
46 bacteria and two archaea \citep{price2024fitnessbrowser}.
For gene $i$, we average its recorded fitness values as
$\bar f_i=|\mathcal{E}_i|^{-1}\sum_{e\in\mathcal{E}_i}f_{ie}$,
where $\mathcal{E}_i$ is the set of recorded experiments,
and assign the binary label
$y_i=\mathbf{1}[\bar f_i\leq-1]$.
All remaining evaluated genes are negatives.
Experiments receive equal weight, including repeated conditions,
and missing values are not imputed.
The target therefore summarizes fitness across the experimental
panel available for each organism.
Within the processed fitness-labeled subset, all genes in an
organism have the same number of recorded experiments,
ranging from 6 to 757 across organisms.

To compute representation sensitivity, \method{} replaces
the target gene's input protein embedding with the organism-wide
mean input embedding, retaining its position, contig, and strand
metadata. This mean is computed within the organism, not across
the pretraining corpus.
Let $\mathcal{N}_i$ contain up to five neighboring genes on either
side of gene $i$ on the same contig, excluding $i$ itself.
With $\vh_j$ and $\vh_j^{(i)}$ denoting the original and perturbed
contextual states, respectively, the sensitivity score is
\begin{equation}
\label{eq:gene-effect-sensitivity}
s_i=
\left\|
\frac{1}{|\mathcal{N}_i|}
\sum_{j\in\mathcal{N}_i}
\left(\vh_j^{(i)}-\vh_j\right)
\right\|_2.
\end{equation}
Neighborhoods are truncated at contig boundaries, and genes
without eligible neighbors are excluded.
We use the canonical 1B checkpoint with 4,096-gene windows
and a 512-gene overlap.

The exploratory Evo2 baseline scores each standalone
forward-strand CDS by its mean autoregressive log-likelihood.
Higher representation sensitivity for \method{} and higher
mean CDS log-likelihood for Evo2 indicate the positive class.
We match model scores to experimental fitness measurements for each gene using its organism and gene identifiers.
Both models are evaluated on the same 179,946 genes with finite
scores and labels.
We compute AUROC separately within each organism and report
the median across organisms.
All 48 organisms contain both classes, with 40--319 positive
genes per organism.

\subsection{Model, Pretraining, and Extraction Details}
\label{app:pretraining-hyperparameters}

Table~\ref{tab:pretraining-hyperparameters} provides model cards for the three scales used in the scaling analysis; the main results use the 1B model.
All models were trained on eight A800 GPUs with an effective batch of 256 genomes.
Metagenome-assembled genomes account for 60\% of training samples, split equally between MGnify and SPIRE (30\% each); the high-quality isolate collection accounts for the remaining 40\%.
We held out 2,000 genomes, evaluated every 5k steps with a fixed 15\% mask (seed 2026), and saved checkpoints every 2.5k steps.
Training and sampling seeds were 42.
For downstream extraction, genomes longer than 4,096 genes are split into overlapping windows with 512-gene overlap, repeated gene states are averaged, and window-level global readouts are averaged.
For genome-level evaluation, Evo2 and NT-v2 process genomic
DNA in non-overlapping windows within each contig.
Evo2 uses 8,192-bp windows and averages the last-token states
from \texttt{blocks.20}, extracted in the forward direction.
NT-v2 uses 6,000-bp windows, averages non-special token states
within each window, and aggregates window representations
weighted by their valid token counts.
Neither pipeline subsamples windows; NT-v2 excludes fragments
shorter than 50 bp.

\begin{table}[h!]
\caption{\method{} architecture and shared pretraining configuration. }
\label{tab:pretraining-hyperparameters}
\begin{center}
\small
\setlength{\tabcolsep}{5pt}
\renewcommand{\arraystretch}{1.03}
\begin{tabularx}{\textwidth}{@{}l*{3}{>{\centering\arraybackslash}X}@{}}
\toprule
Configuration & 85M & 512M & 1B \\
\midrule
\multicolumn{4}{@{}l}{\textbf{Architecture}} \\
\quad Protein representation & \multicolumn{3}{c}{ESM2-35M, 480 dimensions} \\
\quad Encoder hidden size & 768 & 1,280 & 1,408 \\
\quad Encoder layers & 12 & 26 & 44 \\
\quad Encoder attention heads & 12 & 20 & 22 \\
\quad Encoder FFN size & 3,072 & 5,120 & 5,632 \\
\addlinespace[1pt]
\quad Decoder hidden size & 512 & 640 & 640 \\
\quad Decoder layers & 8 & 8 & 8 \\
\quad Decoder attention heads & 16 & 20 & 20 \\
\quad Decoder FFN size & 2,048 & 2,560 & 2,560 \\
\quad Transformer block & \multicolumn{3}{c}{Pre-RMSNorm; GELU; dropout 0.1} \\
\midrule
\multicolumn{4}{@{}l}{\textbf{Genome input and objective}} \\
\quad Genome context & \multicolumn{3}{c}{4,096 genes; 64 global tokens} \\
\quad Genomic structure & \multicolumn{3}{c}{RoPE (random offset $<2{,}048$); contig restriction; strand bias} \\
\quad Masking & \multicolumn{3}{c}{75\% of genes} \\
\quad Reconstruction target & \multicolumn{3}{c}{Global-mean-centered ESM2 embedding} \\
\quad Objective & \multicolumn{3}{c}{$0.3$ MSE $+\,0.7$ cosine} \\
\midrule
\multicolumn{4}{@{}l}{\textbf{Optimization}} \\
\quad Optimizer & \multicolumn{3}{c}{AdamW; $\beta=(0.9,0.95)$; $\epsilon=10^{-8}$; weight decay 0.01} \\
\quad Learning-rate schedule & \multicolumn{3}{c}{$10^{-4}$ peak; 5k-step warmup; cosine decay to $10^{-5}$} \\
\quad Effective batch & \multicolumn{3}{c}{256 genomes on 8 GPUs} \\
\quad Training & \multicolumn{3}{c}{40k steps; BF16; gradient clipping 1.0} \\
\bottomrule
\end{tabularx}
\end{center}
\end{table}
\FloatBarrier

\section{Additional Results}
\label{app:additional-results}

\subsection{Genome-Level Readout Comparison}
\label{app:genome-embedding-geometry}

We compare mean pooling of contextualized gene states with
global-token aggregation using the same 1B mask-75\% checkpoint
at 40k steps.
Both readouts are obtained from the same extraction run and
evaluated with identical species-disjoint splits, labels, and
probe configurations, using five paired probe seeds (1--5).
Table~\ref{tab:genome-readout-comparison} reports the mean
and sample standard deviation using the metrics in
Table~\ref{tab:genome-supervised-results}.

\begin{table}[!htbp]
\centering
\caption{Mean pooling versus global-token aggregation on species-disjoint benchmarks.}
\label{tab:genome-readout-comparison}
\fontsize{8.5}{10}\selectfont
\setlength{\tabcolsep}{3pt}
\renewcommand{\arraystretch}{1.15}
\begin{tabular*}{\linewidth}{@{\extracolsep{\fill}}l*{6}{c}@{}}
\toprule
Readout & Temp. & Doubling & MIC & AMR & \shortstack{Binary\\phen.} & \shortstack{Mixed\\phen.} \\
\midrule
Mean pooling & \shortstack{$0.862$ \\ $\std{0.001}$} & \shortstack{$\mathbf{0.336}$ \\ $\std{0.011}$} & \shortstack{$0.270$ \\ $\std{0.016}$} & \shortstack{$\mathbf{0.682}$ \\ $\std{0.034}$} & \shortstack{$0.794$ \\ $\std{0.004}$} & \shortstack{$\mathbf{0.457}$ \\ $\std{0.027}$} \\
\addlinespace[4pt]
Global tokens & \shortstack{$\mathbf{0.877}$ \\ $\std{0.002}$} & \shortstack{$0.246$ \\ $\std{0.015}$} & \shortstack{$\mathbf{0.281}$ \\ $\std{0.010}$} & \shortstack{$0.672$ \\ $\std{0.019}$} & \shortstack{$\mathbf{0.797}$ \\ $\std{0.003}$} & \shortstack{$0.442$ \\ $\std{0.023}$} \\
\bottomrule
\end{tabular*}
\end{table}

Table~\ref{tab:genome-readout-comparison} shows that neither
readout consistently outperforms the other.
Global-token aggregation performs better on temperature prediction,
whereas mean pooling has a larger advantage on doubling time.
This suggests that the two readouts retain task-relevant information
differently, with no single aggregation strategy preferred across
all evaluated traits.

\FloatBarrier

\FloatBarrier

\subsection{Pretraining Loss Diagnostics}
\label{app:pretraining-loss}

\begin{figure}[!htbp]
    \centering
    \includegraphics[width=\linewidth]{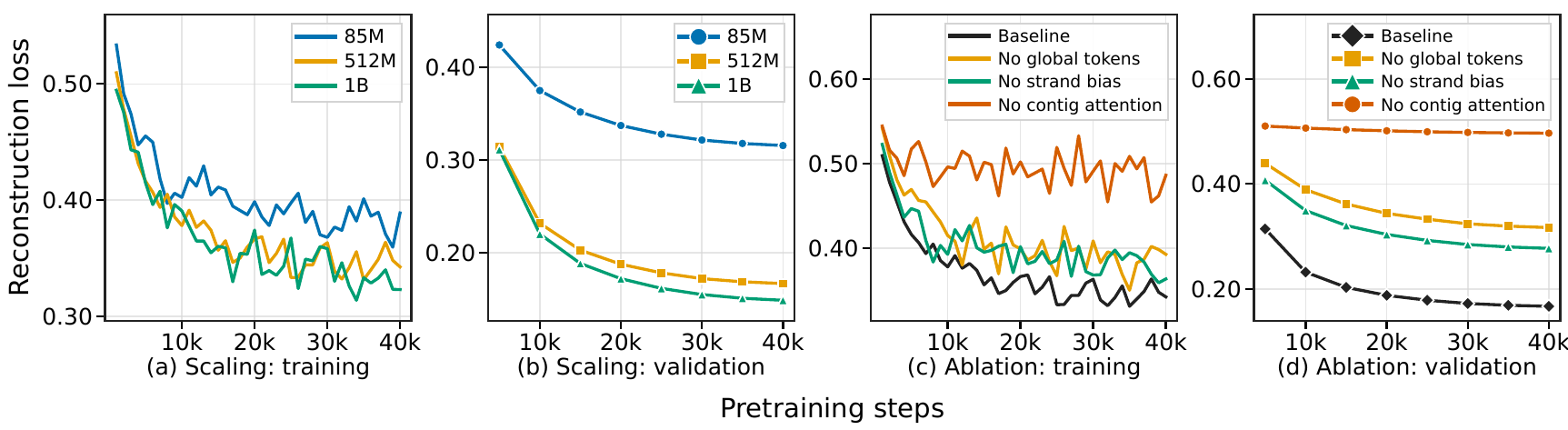}
    \caption{Pretraining reconstruction loss under encoder scaling and architectural ablations. (a,c) Training loss in 1000-step bins. (b,d) Validation loss evaluated every 5k steps with a shared fixed validation mask using 15\% masked genes and seed 2026. Panels (a,b) compare 85M, 512M, and 1B mask-75\% encoders; panels (c,d) compare the 512M mask-75\% baseline with variants that remove global tokens, strand bias, or contig-aware attention.}
    \label{fig:mae-pretraining-loss-diagnostics}
\end{figure}

Figure~\ref{fig:mae-pretraining-loss-diagnostics} compares reconstruction loss under encoder scaling and matched architectural ablations.
Larger encoders reach lower validation loss, consistent with the PPI scaling trend in Figure~\ref{fig:ppi-ablation-summary}(a).
For the 512M model, removing strand bias or global tokens increases validation loss, while removing contig-aware attention causes the largest degradation, mirroring the downstream ordering in Figure~\ref{fig:ppi-ablation-summary}(c).
At 40k steps, validation loss is $0.167$ for the baseline, $0.277$ without strand bias, $0.317$ without global tokens, and $0.497$ without contig-aware attention.

\FloatBarrier

\subsection{Mask Ratio and Pretraining Efficiency}
\label{app:mask-efficiency}

We benchmark the computational cost of pretraining at different gene mask ratios using the 512M model.
Higher masking ratios shorten the encoder input, while the decoder still processes all gene positions.
On a single A800 GPU with batch size one and BF16 precision, we measure forward and backward computation through both the encoder and decoder, including the reconstruction loss, but excluding optimizer updates and distributed communication.
The model, input, and contig-aware attention settings are held fixed.
Reported reductions are averaged over four independent GPU runs, using the median of 20 timed steps after warm-up for runtime and peak allocated memory for memory usage.

\begin{table}[h!]
\caption{Measured pretraining efficiency of 75\% masking (M75) relative to 15\% masking (M15) for the 512M model. $L$ denotes the number of valid genes; F+B denotes forward and backward passes. Encoder token counts include 64 global tokens.}
\label{tab:mask-efficiency-proxy}
\begin{center}
\begin{tabular}{rrrrr}
\toprule
$L$ & \multicolumn{2}{c}{Encoder tokens} & \multicolumn{2}{c}{M75 reduction vs. M15} \\
\cmidrule(lr){2-3}\cmidrule(lr){4-5}
& M15 & M75 & F+B time & Peak memory \\
\midrule
512 & 499 & 192 & 2.5\% & 4.1\% \\
1024 & 934 & 320 & 21.2\% & 27.3\% \\
2048 & 1805 & 576 & 55.9\% & 41.3\% \\
4096 & 3546 & 1088 & 65.2\% & 51.3\% \\
\bottomrule
\end{tabular}
\end{center}
\end{table}

Table~\ref{tab:mask-efficiency-proxy} shows that the measured savings increase with sequence length.
At 4,096 genes, increasing the mask ratio from 15\% to 75\% reduces forward--backward time by $65.2\%$ and peak allocated memory by $51.3\%$.
At 512 genes, runtime decreases by only $2.5\%$ despite the shorter encoder input.
Higher masking ratios thus reduce pretraining runtime and memory
usage, with larger savings for longer genomic inputs in this benchmark.



\end{document}